\documentclass[letterpaper,english,aps,prb,showkeys,twocolumn]{revtex4-1}
\usepackage[T1]{fontenc}
\usepackage[latin9]{inputenc}
\usepackage{array}
\usepackage{booktabs}
\usepackage{units}
\usepackage{textcomp}
\usepackage{amstext}
\usepackage{graphicx}

\makeatletter

\DeclareRobustCommand{\greektext}{%
  \fontencoding{LGR}\selectfont\def\encodingdefault{LGR}}
\DeclareRobustCommand{\textgreek}[1]{\leavevmode{\greektext #1}}
\DeclareFontEncoding{LGR}{}{}
\DeclareTextSymbol{\~}{LGR}{126}
\providecommand{\tabularnewline}{\\}

 \@ifundefined{textcolor}{}
 {%
   \definecolor{BLACK}{gray}{0}
   \definecolor{WHITE}{gray}{1}
   \definecolor{RED}{rgb}{1,0,0}
   \definecolor{GREEN}{rgb}{0,1,0}
   \definecolor{BLUE}{rgb}{0,0,1}
   \definecolor{CYAN}{cmyk}{1,0,0,0}
   \definecolor{MAGENTA}{cmyk}{0,1,0,0}
   \definecolor{YELLOW}{cmyk}{0,0,1,0}
 }

\makeatother

\usepackage{babel}
\begin{document}

\title{Superconductivity in the YIr$_{2}$Si$_{2}$ and LaIr$_{2}$Si$_{2}$
Polymorphs}

\author{Michal Vališka}

\author{Ji\v{r}í Pospíšil}

\author{Jan Prokleška}

\author{Martin Diviš}

\author{Alexandra Rudajevová}

\author{Vladimír Sechovský}

\affiliation{Faculty of Mathematics and Physics, Charles University, DCMP, Ke
Karlovu 5, CZ-12116 Praha 2, Czech Republic}
\begin{abstract}
We report on existence of superconductivity in YIr$_{2}$Si$_{2}$
and LaIr$_{2}$Si$_{2}$ compounds in relation to crystal structure.
The two compounds crystallize in two structural polymorphs, both tetragonal.
The high temperature polymorph (HTP) adopts the CaBe$_{2}$Ge$_{2}$-structure
type (space group P4/nmm) while the low temperature polymorph (LTP)
is of the ThCr$_{2}$Si$_{2}$ type (I4/mmm). By studying polycrystals
prepared by arc melting we have observed that the rapidly cooled samples
retain the HTP even at room temperature (RT) and below. Annealing
such samples at $\geq900^{\circ}\unit{C}$ followed by slow cooling
to RT provides the LTP. Both, the HTP and LTP were subsequently studied
with respect to magnetism and superconductivity by electrical resistivity,
magnetization, AC susceptibility and specific heat measurements. The
HTP and LTP of both compounds respectively, behave as Pauli paramagnets.
Superconductivity has been found exclusively in the HTP of both compounds
below $\mathrm{T_{SC}}$ (= 2.52 K in YIr$_{2}$Si$_{2}$ and 1.24
K in LaIr$_{2}$Si$_{2}$). The relations of magnetism and superconductivity
with the electronic and crystal structure are discussed with comparing
experimental data with the results of first principles electronic
structure calculations.
\end{abstract}


\maketitle

\section{introduction}

To date large variety of ternary intermetallic systems has been discovered
and many of them remain subjects of intensive scientific interest.
The rare-earth intermetallics are the most intensively studied with
respect to the cooperative phenomena \textendash{} magnetism and superconductivity.
The localized 4\emph{f} electrons of the rare-earth ions exposed to
various crystallographic and chemical environments in compounds cause
a rich spectrum of the physical phenomena like magnetism with various
magnetic structures, heavy fermion behavior or unconventional superconductivity
in the quantum critical regime. The weak interactions of localized
4\emph{f} electrons with valence electrons of neighboring ions, which
are the key ingredients of the physics of these intermetallics, are
strongly dependent on the complex interplay of chemistry and symmetry
of the neighborhood. Studies of La, Lu or Y analogues allow inspecting
properties independent on presence of 4\emph{f}-electrons in the material.

Only in some specific cases one can investigate the effect of crystal
symmetry change alone without changing composition. An outstanding
opportunity is offered by polymorphism. Polymorphism is a unique phenomenon
when the material can appear in two or even few different crystal
structures although the chemical stoichiometry and composition are
conserved. It means that the same collection of ions can be arranged
to two or more crystal structures of different symmetry.

One large group of materials exhibiting often polymorphism is characterized
by the composition \emph{$RET_{2}X_{2}$}, where $RE$ represents
various rare-earth ions, $T$ stays for 3\emph{d}, 4\emph{d} and 5\emph{d}
transition metals and noble metals, e.g. Ag or Au and \emph{X} represents
\emph{p}-elements like Si , Ge, As or P\cite{45,50,51,15,32,33,34,10,11,12,16,46,5,40}.
Two polymorphs can be found in the case of \emph{122} iridium silicides.
One polymorph crystallizes in the primitive tetragonal structure of
the CaBe$_{2}$Ge$_{2}$ type (space group P4/nmm) and the second
adopts the ThCr$_{2}$Si$_{2}$-type body centered tetragonal structure
(I4/mmm). The ThCr$_{2}$Si$_{2}$\textendash{}type is thermodynamically
stable at room temperature (LTP) while the CaBe$_{2}$Ge$_{2}$-type
is thermodynamically stable at high temperatures (HTP). Nevertheless,
the HTP can exist as a metastable form at room temperature when the
cooling rate from the melt is high enough. The temperatures of the
structural transitions between polymorphs are the characteristic parameters
of each polymorphic compound. Existence of the polymorphism is probably
not the general rule in the entire group of the $RE$Ir$_{2}$Si$_{2}$
compounds. For example EuIr$_{2}$Si$_{2}$ is reported crystallizing
only in the CaBe$_{2}$Ge$_{2}$-type structure\cite{15,41}.

The group of \emph{122} rare-earth-iridium silicides exhibits variety
of phenomena like magnetism, superconductivity or non-Fermi liquid
behavior. A short overview is presented in the following paragraph
together with motivation of our work.

The Ce analogue does not order magnetically, neither the HTP nor the
LTP variant presumably due to lack of Ce magnetic moments of valence
fluctuating Ce ions. The LTP behaves as the Fermi-liquid at low temperatures
whereas the HTP exhibits non-Fermi-liquid features\cite{42,6}. It
is in contrast to the uranium polymorphs, which order antiferromagneticaly
in the LTP whereas the HTP is paramagnetic at low temperatures\cite{45,46,47}.

Similar situation to U compounds has been found in the Pr polymorphs.
The HTP is paramagnetic down to 2 K while the LTP orders antiferromagneticaly
at $T_{\mathrm{N}}=\unit[45.5]{K}$. Magnetism of the Pr polymorphs
is determined by the crystal field (CF) acting on the Pr ion\cite{44,43,14,2}.
The Nd compounds behave likewise the Pr ones. The HTP is entirely
paramagnetic whereas the LTP becomes antiferromagnetic below $T_{\mathrm{N}}=\unit[33]{K}$\cite{14,104}.
Existence of the SmIr$_{2}$Si$_{2}$ polymorphs has been confirmed
by a structure study but information regarding magnetism is still
missing\cite{39}. As already mentioned EuIr$_{2}$Si$_{2}$ has been
reported crystallizing only in the CaBe$_{2}$Ge$_{2}$ type structure
and exhibits intermediate valence behavior\cite{41,39}. Rather incomplete
information is available for the heavy rare-earth iridium silicides;
for TmIr$_{2}$Si$_{2}$ no scientific data at all are available and
for HoIr$_{2}$Si$_{2}$ besides confirmation of polymorphism\cite{39}
no other information has been published. The LTP of Gd, Tb, Dy and
Er are reported to be antiferromagnetic below $T_{\mathrm{N}}=80,80,40$
and $\unit[10]{K}$, respectively \cite{50,104,39,55,57,54,56}. Antiferromagnetism
($T_{\mathrm{N}}=\unit[13]{K}$) is reported also for the HTP of TbIr$_{2}$Si$_{2}$\cite{55};
no other physical properties of HTP are known for the heavy rare-earth
iridium silicides except for the Yb compound. YbIr$_{2}$Si$_{2}$
attracted much interest owing to heavy fermion behavior of the LTP,
which is close to a quantum critical point (QCP)\cite{55,57}. In
contrary the HTP orders magnetically at 0.7 K\cite{55,54,56}. No
information on physical properties of LuIr$_{2}$Si$_{2}$ has been
published although the Lu compound was often mentioned as a non-magnetic
analogue within the YbIr$_{2}$Si$_{2}$ study.

Superconductivity has been observed for the HTP of two $RE$Ir$_{2}$Si$_{2}$
compounds without the 4\emph{f} electrons, namely YIr$_{2}$Si$_{2}$
and LaIr$_{2}$Si$_{2}$ . The literature reports, which are briefly
reviewed in Table \ref{tab:Tsc}, are in some cases contradictory,
especially as concerns presence of superconductivity in the LTP of
YIr$_{2}$Si$_{2}$. In this context we would like to note that lanthanum
at ambient pressure becomes a superconductor below about 6 K in both
the h.c.p and f.c.c. crystal form, respectively in ambient pressure\cite{106}.
In contrary yttrium metal under ambient pressure shows no superconductivity
but Wittig\cite{107} has discovered superconductivity in yttrium
metal at 1.2 K when applying pressure of 11 GPa. 
\begin{table}
\begin{tabular}{ccc}
\toprule 
LaIr$_{2}$Si$_{2}$  & HTP $T_{\mathrm{SC}}$(K) & LTP $T_{\mathrm{SC}}$(K)\tabularnewline
\midrule
\midrule 
Ref\cite{1} & 1.56 & -\tabularnewline
Ref\cite{4} & 1.52-1.58 & -\tabularnewline
Ref\cite{30} & 1.6 & -\tabularnewline
\midrule
\midrule 
YIr$_{2}$Si$_{2}$  & HTP $T_{\mathrm{SC}}$(K) & LTP $T_{\mathrm{SC}}$(K)\tabularnewline
\midrule
\midrule 
Ref\cite{4} & 2.72-2.83 & -\tabularnewline
Ref\cite{13} & 2.6 & 2.6 (broaden)\tabularnewline
Ref\cite{18} & 2.7 & 2.4\tabularnewline
Ref\cite{19} & 2.7 & 2.45\tabularnewline
\bottomrule
\end{tabular}

\caption{\label{tab:Tsc}The overview of the presence and temperatures of superconducting
transitions in Y and La iridium silicides. }

\end{table}

The primary objective of the present paper is to clarify how magnetism
and superconductivity in YIr$_{2}$Si$_{2}$  and LaIr$_{2}$Si$_{2}$
 is connected with the specific layered crystal structures of the
two polymorphs (LTP and HTP). As a first step we studied existence
of the LTP- and HTP-phase, respectively, with respect to thermal history
by combining the DTA measurements with X-ray powder diffraction analysis
of samples at room temperature. The main part of the paper is devoted
to the low temperature measurements of the electrical resistivity,
magnetization, AC susceptibility and specific heat in various magnetic
fields. To corroborate our explanation of experimental results we
have also performed \emph{ab initio} electronic structure calculations
for YIr$_{2}$Si$_{2}$  and theoretical predictions regarding the
superconducting state.

\section{experimental and computation details}

In order to avoid problems with stabilizing the HTP at low temperatures
when growing the single crystals by Czochralski method\cite{104}
we have decided to perform the work on polycrystals which can be easier
quenched after melting and single-HTP samples can be obtained. The
samples of YIr$_{2}$Si$_{2}$  and LaIr$_{2}$Si$_{2}$  compounds
have been prepared by melting the stoichiometric amounts of elements
(purity of Y and La - 99.9\%, Ir - 99.99\%, Si - 99.9999\%) in an
arc-furnace with a water-cooled copper crucible under the high-purity
(6N) argon protective atmosphere. The total sample mass was typically
2.5 g. The samples were re-melted three times to ensure good homogeneity;
finally the sample has been left to cool rapidly after sudden switching
off the arc above the melt. No significant evaporation has been observed
during melting the samples. Each sample has been cut into two equal
parts. One half of the each sample was wrapped in a tantalum foil
(99.9 \%), sealed in a quartz tube under the vacuum of $\unit[1\cdot10^{-6}]{mbar}$,
annealed at $900^{\circ}\unit{C}$ for 7 days and then slowly cooled
to avoid internal stresses. The second half was kept without any heat
treatment. Both the samples (as cast and annealed) were characterized
by the X-ray powder diffraction method (XRPD) at room temperature
using a Bruker D8 Advance diffractometer equipped with a monochromator
providing the CuK$\alpha$ radiation. The diffraction patterns were
evaluated by the standard Rietveld technique\cite{63} using the FullProf/WinPlotr
software\cite{102}. The composition of samples was verified by chemical
analysis using a scanning electron microscope (SEM) Tescan Mira I
LMH equipped by an energy dispersive X-ray detector (EDX) Bruker AXS.
The samples have been shaped appropriate for individual measurements
using a fine wire saw to prevent additional stresses in the samples.

Differential thermal analysis (DTA) measurements were performed using
Setaram SETSYS-2400 CS instrument over the range from room temperature
to $1450^{\circ}\unit{C}$. The heating and cooling rates were 5 K/min.

The low temperature behavior was tested by the electrical resistivity,
heat capacity, magnetization and AC susceptibility measurements in
the PPMS9T and MPMS7T facilities (Quantum Design). The samples for
the electrical resistivity measurement were of the bar shape (1 mm
x 1 mm x 4 mm). The electrical resistivity was measured as a function
of temperature and magnetic field by using the four-terminal AC method.
The heat capacity was measured on the 1.5 mm x 1.5 mm x 0.5 mm plates
by the relaxation method. The magnetic measurements were performed
with fine powder samples having the grains fixed in random orientation
by weakly magnetic glue. The electrical resistivity was measured also
with a $^{3}$He option down to the temperature of 350 mK\cite{109}.

The results of calculations of the electronic structure of LaIr$_{2}$Si$_{2}$
have been published before\cite{2}. The ground-state electronic structure
YIr$_{2}$Si$_{2}$ was calculated on the basis of DFT within the
local spin density approximation (LSDA)\cite{61} and the generalized
gradient (GGA) approximation\cite{64,62,99} therefore we used more
possibilities than just GGA\cite{62} in the work\cite{100}. For
this purpose, we used the full-potential augmented-plane-wave plus
local-orbitals method (APW-lo) as implemented in the latest version
(WIEN2k) of the original WIEN code\cite{101}. The calculations were
performed within scalar relativistic mode with the following parameters.
Non-overlapping atomic-sphere (AS) radii of 2.8, 2.3 and 1.6 a.u.
(1 a.u. = 52.9177 pm) were taken for Y, Ir and Si, respectively. The
basis for the expansion of the valence states (less than 8 Ry below
the Fermi energy) consisted of more than 3500 basis functions (more
than 350 APW/ atom) for the HTP-structure and more than 1350 (more
than 270 APW/atom) for the LTP-structure plus the Y 3\emph{s}, 3\emph{p},
Ir 5\emph{s}, 5\emph{p}, 4\emph{f} and Si 2\emph{p} local orbitals.
The Brillouin-zone integrations were performed with the tetrahedron
method 47, on a 330 k-point mesh (HTP) and 288 (LTP) corresponding
to more than 4000 k-points in the full Brillouin zone for both the
crystal structures.

\section{results and discussion}

\subsection{Composition and crystal structure analysis}

We have successfully synthesized the polycrystalline samples of the
YIr$_{2}$Si$_{2}$ and LaIr$_{2}$Si$_{2}$ compounds and performed
FE-EDX analysis at various locations of the annealed samples. The
element analysis confirmed the majority of the YIr$_{2}$Si$_{2}$
and LaIr$_{2}$Si$_{2}$ phases, respectively, in all surveyed samples
and 1-2 percent of the La rich and Si poor phase (La$_{1.04}$Ir$_{2}$Si$_{1.58}$)
localized mainly on grain boundaries of the La based samples.

\begin{figure}
\includegraphics{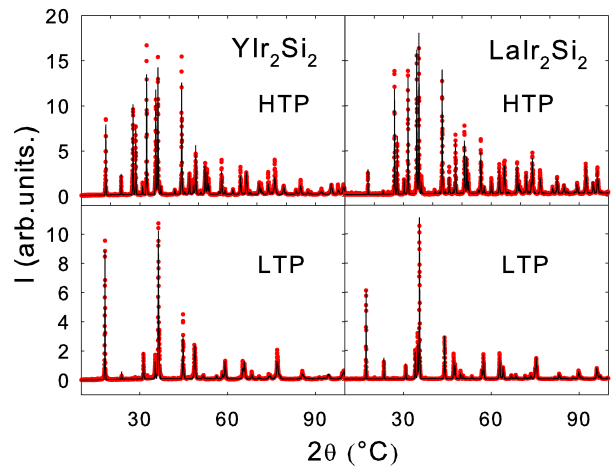}

\caption{\label{fig:XRPD1}XRPD patterns of the YIr$_{2}$Si$_{2}$ and LaIr$_{2}$Si$_{2}$
compounds. The upper panels represent the patterns of the as-cast
samples. The lower panels represent the patterns of the annealed materials
as described in the text.}

\end{figure}

Subsequently, small pieces of the as-cast and annealed samples of
both compounds, were powdered in agate mortar and XRPD data have been
collected. We have observed significant difference between the patterns
of the as-cast and annealed samples in the case of both compounds
due to the crystal structure transformation within the thermal treatment.
All reflections can be assigned to the ThCr$_{2}$Si$_{2}$\textendash{}type
structure type in the case on thermally treated samples while the
as-cast samples kept the CaBa$_{2}$Ge$_{2}$-structure type. No significant
additional reflections have been found in either sample. The powder
patterns of the both compounds in the as-cast and annealed states
are displayed in Figure \ref{fig:XRPD1}.

The proper analysis of the XRPD pattern of the YIr$_{2}$Si$_{2}$
HTP revealed a mixture between the Si and Ir atoms in the 2\emph{c}
positions. The refinement process showed mixture of about 10 \textendash{}
12 \%. A similar situation has been observed in the case of LaIr$_{2}$Si$_{2}$
with the Si/Ir mixture of about 5 \%. On the other hand no atoms mixture
has been detected in the LTP and all atoms of the involved elements
well occupied their positions in both compounds. The crystal structure
parameters of the YIr$_{2}$Si$_{2}$ compound are digestedly summarized
in the Table \ref{tab:Y_HTP} and Table \ref{tab:Y_LTP}. The determined
parameters are in good agreement with the structure model published
by Shelton et al.\cite{4}.

\begin{table}
\begin{tabular}{ccccc}
\toprule 
\multicolumn{3}{c}{YIr$_{2}$Si$_{2}$ \textendash{} HTP} & $a\,(\textrm{\AA})$ & 4.0938(2)\tabularnewline
\multicolumn{3}{c}{CaBa$_{2}$Ge$_{2}$ structure type} & $c\,(\textrm{\AA})$ & 9.6856(7)\tabularnewline
\midrule
\midrule 
Atoms & Symmetry & x & y & z\tabularnewline
\midrule
\midrule 
Ir & 2\emph{c} & \textonequarter{} & \textonequarter{} & 0.127(5)\tabularnewline
Si & 2\emph{c} & \textonequarter{} & \textonequarter{} & 0.422(9)\tabularnewline
Y & 2\emph{c} & \textonequarter{} & \textonequarter{} & 0.742(8)\tabularnewline
Ir & 2\emph{b} & \textthreequarters{} & \textonequarter{} & \textonehalf{}\tabularnewline
Si & 2\emph{a} & \textthreequarters{} & \textonequarter{} & 0\tabularnewline
\bottomrule
\end{tabular}

\caption{\label{tab:Y_HTP}Crystal structure parameters of the HTP of YIr$_{2}$Si$_{2}$.}

\end{table}

\begin{table}
\begin{tabular}{ccccc}
\toprule 
\multicolumn{3}{c}{YIr$_{2}$Si$_{2}$ \textendash{} LTP} & $a\,(\textrm{\AA})$ & 4.0483(2)\tabularnewline
\multicolumn{3}{c}{ThCr$_{2}$Si$_{2}$ structure type} & $c\,(\textrm{\AA})$ & 9.8152(7)\tabularnewline
\midrule
\midrule 
Atoms & Symmetry & x & y & z\tabularnewline
\midrule
\midrule 
Si & 4e & 0 & 0 & 0.3640(9)\tabularnewline
Ir & 4d & 0 & \textonehalf{} & \textonequarter{}\tabularnewline
Y & 2a & 0 & 0 & 0\tabularnewline
\bottomrule
\end{tabular}

\caption{\label{tab:Y_LTP}Crystal structure parameters of the LTP of YIr$_{2}$Si$_{2}$.}

\end{table}

Lattice parameters and fractional coordinates of the both LaIr$_{2}$Si$_{2}$
polymorphs are summarized in the Table \ref{tab:La_HTP} and Table
\ref{tab:La_LTP}.

\begin{table}
\begin{tabular}{ccccc}
\toprule 
\multicolumn{3}{c}{LaIr$_{2}$Si$_{2}$ \textendash{} HTP} & $a\,(\textrm{\AA})$ & 4.1873(1)\tabularnewline
\multicolumn{3}{c}{CaBa$_{2}$Ge$_{2}$ structure type} & $c\,(\textrm{\AA})$ & 9.9380(3)\tabularnewline
\midrule
\midrule 
Atoms & Symmetry & x & y & z\tabularnewline
\midrule
\midrule 
Ir & 2c & \textonequarter{} & \textonequarter{} & 0.374(4)\tabularnewline
Si & 2c & \textonequarter{} & \textonequarter{} & 0.140(8)\tabularnewline
La & 2c & \textonequarter{} & \textonequarter{} & 0.742(8)\tabularnewline
Ir & 2a & \textthreequarters{} & \textonequarter{} & 0\tabularnewline
Si & 2b & \textthreequarters{} & \textonequarter{} & \textonehalf{} \tabularnewline
\bottomrule
\end{tabular}

\caption{\label{tab:La_HTP}Crystal structure parameters of the HTP of LaIr$_{2}$Si$_{2}$.}

\end{table}

\begin{table}
\begin{tabular}{ccccc}
\toprule 
\multicolumn{3}{c}{LaIr$_{2}$Si$_{2}$ \textendash{} LTP} & $a\,(\textrm{\AA})$ & 4.1111(1)\tabularnewline
\multicolumn{3}{c}{ThCr$_{2}$Si$_{2}$ structure type} & $c\,(\textrm{\AA})$ & 10.3001(3)\tabularnewline
\midrule
\midrule 
Atoms & Symmetry & x & y & z\tabularnewline
\midrule
\midrule 
Si & 4e & 0 & 0 & 0.366(1)\tabularnewline
Ir & 4d & 0 & \textonehalf{} & \textonequarter{}\tabularnewline
La & 2a & 0 & 0 & 0\tabularnewline
\bottomrule
\end{tabular}

\caption{\label{tab:La_LTP}Crystal structure parameters of the LTP of LaIr$_{2}$Si$_{2}$.}

\end{table}

The difference between corresponding lattice parameters of LaIr$_{2}$Si$_{2}$
and YIr$_{2}$Si$_{2}$ compounds is about 1 \% which well corresponds
with the difference between the La and Y atomic radii. The \emph{c/a}
ratio of HTP is 1\% and 3\% larger than the value for the LTP of YIr$_{2}$Si$_{2}$
and LaIr$_{2}$Si$_{2}$, respectively.

\begin{figure}
\includegraphics{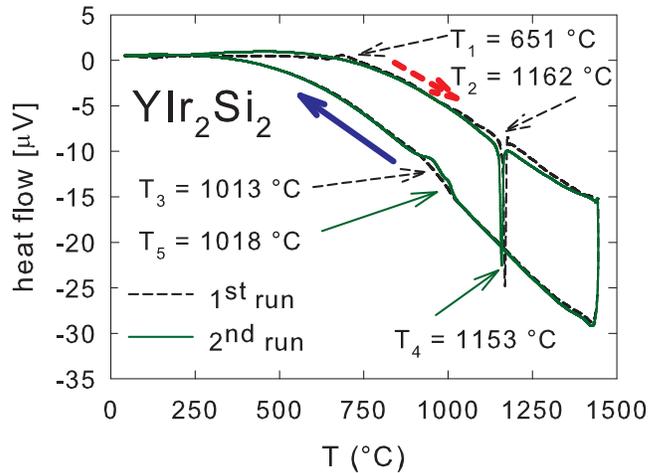}

\caption{\label{fig:DTA}DTA curves recorded for YIr$_{2}$Si$_{2}$. The upper
red bold (dashed) arrow denotes the heating branch. The lower blue
bold (solid) arrow denotes the cooling branch. Temperatures of transitions
are numbered and marked by black (dashed) arrows for first run and
green (solid) arrow for second run. }

\end{figure}

The transformation temperatures between the HTP and LTP in both compounds
were studied by DTA and the typical result for the YIr$_{2}$Si$_{2}$
compound is displayed in Figure \ref{fig:DTA}. Two thermal cycles
have been applied on the as cast samples, which contain entirely the
HTP. Two anomalies appeared in the first heating branch of the both
compounds. When heating from room temperature the first anomaly corresponds
with the transformation from the HTP, which is metastable at room
temperature to the thermodynamically preferable LTP. The transformation
temperature has been found $\sim365^{\circ}\unit{C}$ ($651^{\circ}\unit{C}$)
for the LaIr$_{2}$Si$_{2}$ (YIr$_{2}$Si$_{2}$) compound. It is
in good agreement with transition temperature $340^{\circ}\unit{C}$
for LaIr$_{2}$Si$_{2}$ compound reported by Mihalik et al.\cite{108}.
The second anomaly appears when the LTP transforms to the HTP at $1101^{\circ}\unit{C}$
for LaIr$_{2}$Si$_{2}$ (\textasciitilde{}$1140^{\circ}\unit{C}$
by Mihalik et al.\cite{108}) and $1162^{\circ}\unit{C}$ for YIr$_{2}$Si$_{2}$.
Only one transition has been found on the cooling branch, which corresponds
to the reverse transition from the HTP to the LTP at $800^{\circ}\unit{C}$
for the LaIr$_{2}$Si$_{2}$ (\textasciitilde{}$860^{\circ}\unit{C}$
by Mihalik et al.\cite{108}) and $1013^{\circ}\unit{C}$ for the
YIr$_{2}$Si$_{2}$ compound (see Figure \ref{fig:DTA}). 

In the case of YIr$_{2}$Si$_{2}$ the sample after the first thermal
cycle can be considered as being in the well-annealed state, i.e.
containing only the LTP. Therefore the 651$^{\circ}\unit{C}$ anomaly
is obviously missing on the heating branch of the second thermal cycle
whereas the other features well correspond to these observed in the
first thermal cycle.

The observed transformation temperatures are digestedly summarized
in the Table \ref{tab:DTA}.

\begin{table}
\begin{tabular}{cccc}
\toprule 
Transf. T ($^{\circ}\unit{C}$) & YIr$_{2}$Si$_{2}$ & LaIr$_{2}$Si$_{2}$ & LaIr$_{2}$Si$_{2}$\cite{108} \tabularnewline
\midrule
\midrule 
$T_{1}$ & 651 & 365 & 340\tabularnewline
$T_{2}$ & 1162 & 1101 & 1140\tabularnewline
$T_{3}$ & 1013 & 800 & 860\tabularnewline
$T_{4}$ & 1153 & 1107 & not measured\tabularnewline
$T_{5}$ & 1018 & -  & not measured\tabularnewline
$\left|T_{2}-T_{3}\right|$ & 149 & 301 & not measured\tabularnewline
\bottomrule
\end{tabular}

\caption{\label{tab:DTA}List of transition temperatures for all compounds,
First run: $T_{1}$ - temperature of the transformation of the metastable
(at room temperature) HTP to stable LTP, $T_{2}$ - temperature of
the transition from the stable LTP to the stable HTP, $T_{3}$ - temperature
of the transition from the stable HTP to the stable LTP; second run:
$T_{4}$ - temperature of the transition from the stable LTP to the
stable HTP, $T_{5}$ - temperature of the transition from the stable
HTP to the stable LTP, $\left|T_{2}-T_{3}\right|$represents thermal
hysteresis of HTP-LTP transformation in the $1^{\mathrm{st}}$run.}
\end{table}

To achieve the desirable YIr$_{2}$Si$_{2}$ and LaIr$_{2}$Si$_{2}$
samples for measurements of low temperature properties which contain
entirely the LPT we annealed the as cast samples in the temperature
interval $800-900^{\circ}\unit{C}$ (which is high enough to transform
the frozen HTP to the LTP but low enough to prevent the LTP to the
HTP transformation. The cooling rate used after annealing was lower
than 1 K/min. Then the annealed samples have been found free of any
traces of HTP.

\subsection{Resistivity and heat capacity results}

\begin{figure}
\includegraphics{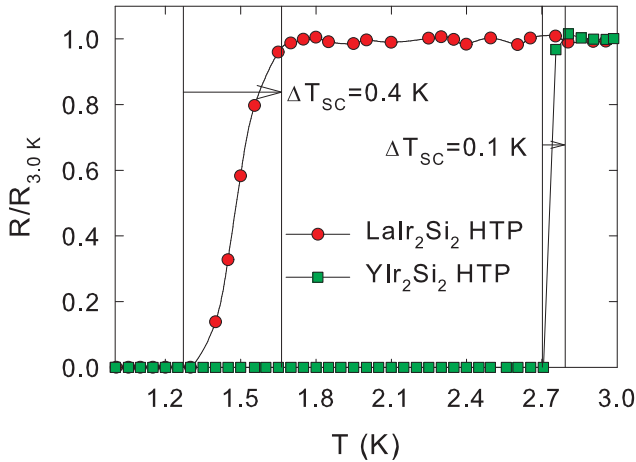}

\caption{\label{fig:ACT_both}Temperature dependence of the electrical resistivity
for the LaIr$_{2}$Si$_{2}$ HTP and YIr$_{2}$Si$_{2}$ HTP. The
sharp drop of resistivity signs the superconductivity transition.}

\end{figure}

We have measured electrical resistivity data of the as-cast (HTP)
and annealed (LTP) samples of both compounds and we have observed
significantly different behavior. While the HTP of both compounds,
respectively, exhibit robust superconductivity \textendash{} see the
drop of the resistivity around the superconducting transition temperature
$T_{\mathrm{SC}}$, which has been assigned to the maximum slope of
the resistivity drop (Figure \ref{fig:ACT_both}). We have found the
values of $T_{\mathrm{SC}}=\unit[2.72]{K}$ for YIr$_{2}$Si$_{2}$
and $T_{\mathrm{SC}}=\unit[1.4]{K}$ for LaIr$_{2}$Si$_{2}$. Nevertheless
there is a significant difference between the SC transition of the
La and Y compound. While a very sharp transition from normal to superconducting
state within 0.1 K has been detected for the Y compound a broader
transition (0.4 K) occurs in the La compound (see Figure \ref{fig:ACT_both}).
The relatively high temperatures of the superconducting transition
in HTP (higher than 1 K) are attributed to the unique arrangement
of the Ir atoms in the 5 Si pyramidal cages in the HTP crystal structure.
It is therefore tempting to speculate that the transition-metals coordination
in these silicides is linked with and may be crucial for the occurrence
of superconductivity in these compounds\cite{30}, which may happen
if the material is not well annealed. No sign of superconductivity
has been detected in the LTP polymorphs down to the lowest temperature
of our experiment (0.4K). It is in contradiction with some previous
papers where superconductivity in YIr$_{2}$Si$_{2}$ was reported
for both the HTP and LTP samples\cite{13,19}. We believe that the
main reason of superconductivity observed in the LTP samples reported
in some earlier papers can be attributed to the presence of certain
amount of the HTP in these samples.

\begin{figure}
\includegraphics{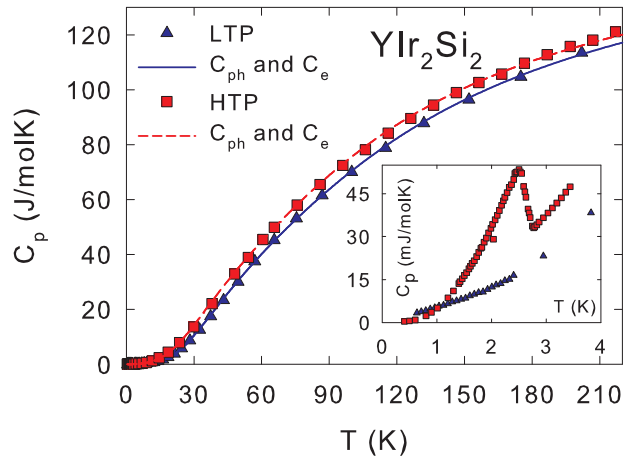}

\caption{\label{fig:HC Y}Temperature dependence of heat capacity of the YIr$_{2}$Si$_{2}$
sample. The marks represent measured data where every second point
has been omitted for clarity. The lines correspond to the fitted models.}

\end{figure}

\begin{figure}
\includegraphics{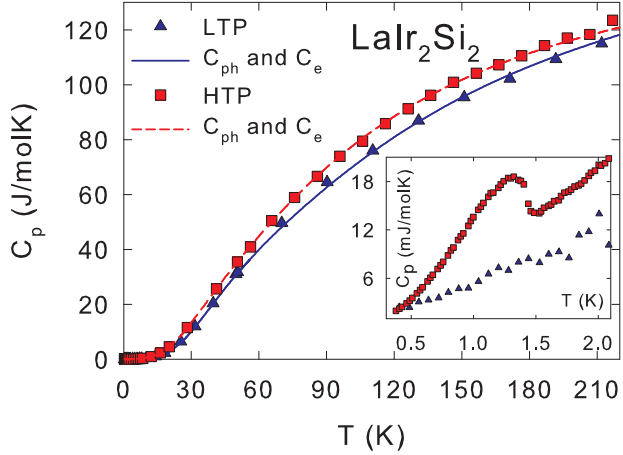}

\caption{\label{fig:HC La}Temperature dependence of heat capacity of the LaIr$_{2}$Si$_{2}$
sample. The marks represent measured data where every second point
has been omitted for clarity. The lines correspond to the fitted models.}

\end{figure}

To confirm the bulk superconducting properties we have measured the
heat capacity data of all four samples and have observed also here
significantly different behavior of the HTP and LTP samples of corresponding
compounds similar to resistivity behaviour. While the HTP of both
compounds were superconducting \textendash{} see the anomaly at temperatures
of the superconducting transitions in the heat capacity data (Figure
\ref{fig:HC Y} and Figure \ref{fig:HC La}) no sign of any transition
has been observed in the data of the LTPs. The $T_{\mathrm{SC}}$
values determined from the heat capacity behavior are somewhat lower
than these indicated by resistivity data. We have found the critical
values based on heat capacity data of $T_{\mathrm{SC}}=\unit[2.52]{K}$
for the YIr$_{2}$Si$_{2}$ and $T_{\mathrm{SC}}=\unit[1.24]{K}$
for the LaIr$_{2}$Si$_{2}$ compound. The $T_{\mathrm{SC}}$ values
of the both HTP of both compounds well correspond with values found
for HTP in literature which are listed in the Table \ref{tab:Tsc}.

We have analyzed the heat capacity considering it as a sum of electron
($C_{\mathrm{e}}$) and phonon ($C_{\mathrm{ph}}$) contribution: 

\begin{equation}
C_{\mathrm{e}}=\frac{2nk_{\mathrm{B}}^{2}T}{E_{\mathrm{F}}}=\gamma T
\end{equation}

where $n$ is the density of electronic states at Fermi level $E_{\mathrm{F}}$
and $k_{\mathrm{B}}$ is Boltzmann constant.

The phonon contribution was treated within the Debye and Einstein
models using anharmonicity correction\cite{105}. The values of the
Debye and Einstein temperatures and degeneracy of the Einstein modes
are listed in Table \ref{tab:Phonon_Y} and Table \ref{tab:Phonon_La}.

\begin{table}
\begin{tabular}{ccc>{\centering}m{1.5cm}cc>{\centering}m{1.5cm}}
\toprule 
YIr$_{2}$Si$_{2}$ & \multicolumn{3}{c}{HTP} & \multicolumn{3}{c}{LTP}\tabularnewline
\midrule
\midrule 
Branch & Degen. & $\theta\unit{\left[K\right]}$ & $\alpha\cdot10^{-4}$\linebreak$\unit{\left[K^{-1}\right]}$ & Degen. & $\theta\unit{\left[K\right]}$ & $\alpha\cdot10^{-4}$\linebreak$\unit{\left[K^{-1}\right]}$\tabularnewline
$\theta_{\mathrm{D}}$ &  & 145 &  &  & 180 & 4\tabularnewline
$\theta_{\mathrm{E1}}$ & 2 & 170 & 4 & 2 & 180 & 4\tabularnewline
$\theta_{\mathrm{E2}}$ & 3 & 180 & 4 & 3 & 185 & 4\tabularnewline
$\theta_{\mathrm{E3}}$ & 3 & 405 & 4 & 3 & 390 & 4\tabularnewline
$\theta_{\mathrm{E4}}$ & 4 & 410 & 4 & 4 & 395 & 4\tabularnewline
\bottomrule
\end{tabular}

\caption{\label{tab:Phonon_Y}Phonon spectrum of the both polymorphs of the
YIr$_{2}$Si$_{2}$ compound.}

\end{table}

\begin{table}
\begin{tabular}{ccc>{\centering}m{1.5cm}cc>{\centering}m{1.5cm}}
\toprule 
LaIr$_{2}$Si$_{2}$ & \multicolumn{3}{c}{HTP} & \multicolumn{3}{c}{LTP}\tabularnewline
\midrule
\midrule 
Branch & Degen. & $\theta\unit{\left[K\right]}$ & $\alpha\cdot10^{-4}$\linebreak$\unit{\left[K^{-1}\right]}$ & Degen. & $\theta\unit{\left[K\right]}$ & $\alpha\cdot10^{-4}$\linebreak$\unit{\left[K^{-1}\right]}$\tabularnewline
$\theta_{\mathrm{D}}$ &  & 155 &  &  & 200 & 5\tabularnewline
$\theta_{\mathrm{E1}}$ & 2 & 170 & 4 & 2 & 170 & 5\tabularnewline
$\theta_{\mathrm{E2}}$ & 3 & 175 & 4 & 3 & 175 & 5\tabularnewline
$\theta_{\mathrm{E3}}$ & 3 & 390 & 4 & 3 & 470 & 5\tabularnewline
$\theta_{\mathrm{E4}}$ & 4 & 395 & 4 & 4 & 475 & 5\tabularnewline
\bottomrule
\end{tabular}

\caption{\label{tab:Phonon_La}Phonon spectrum of the both polymorphs of the
LaIr$_{2}$Si$_{2}$ compound.}
\end{table}

Values of the Sommerfeld \textgreek{g} coefficient determined from
the specific heat data are summarized in Table \ref{tab:Sommerfeld}.

\begin{table}
\begin{tabular}{ccccc}
\toprule 
 & \multicolumn{2}{c}{YIr$_{2}$Si$_{2}$} & \multicolumn{2}{c}{LaIr$_{2}$Si$_{2}$}\tabularnewline
\cmidrule{2-5} 
 & HTP & LTP & HTP & LTP\tabularnewline
\midrule
\midrule 
$\unit[\gamma]{\left[mJ\cdot mol^{-1}\cdot K^{-2}\right]}$ & 8 & 8 & 8.1 & 8.1\tabularnewline
\bottomrule
\end{tabular}

\caption{\label{tab:Sommerfeld}Sommerfeld $\gamma$ coefficient of all compounds.}

\end{table}

Our experimental values of $\gamma$ coefficient can compare as well
with the coefficients presented by Braun et al.\cite{1}. The value
of $\unit[\gamma=8.5]{mJ\cdot mol^{-1}\cdot K^{-2}}$ presented by
Braun et al. for HTP is in good agreement with our observed value.
On the other hand our experimental value of $\gamma$ is significantly
higher than the value $\unit[\gamma=4.5]{mJ\cdot mol^{-1}\cdot K^{-2}}$
presented by Braun et al.\cite{1} for the LTP.

\begin{figure}
\includegraphics{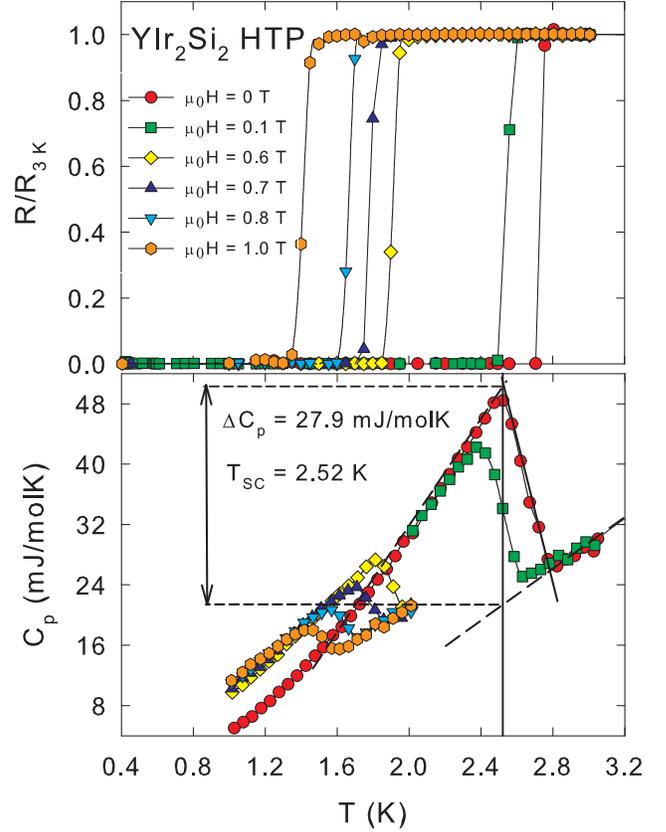}

\caption{\label{fig:HC+ACT_Y}Temperature dependence of the specific heat and
electric resistivity of the HTP of YIr$_{2}$Si$_{2}$ measured in
various magnetic fields.}

\end{figure}

\begin{figure}
\includegraphics{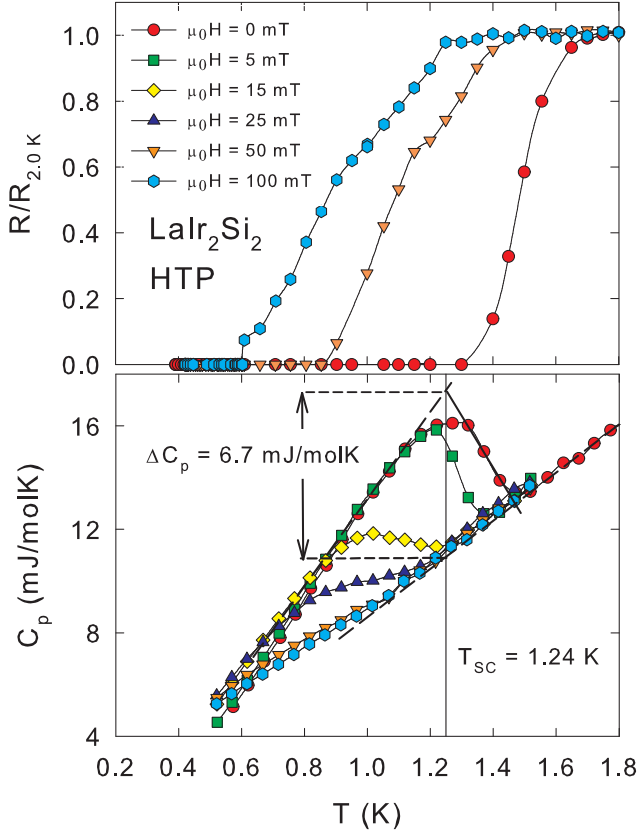}

\caption{\label{fig:ACT+HC_La}Temperature dependence of the specific heat
and electric resistivity of the HTP of LaIr$_{2}$Si$_{2}$ measured
in various magnetic fields.}

\end{figure}

In order to analyze the superconducting properties of both compounds
with respect to the predictions of BCS theory we have measured the
heat capacity and electrical resistivity at low temperatures under
various magnetic fields. The results are shown in Figure \ref{fig:HC+ACT_Y}
and \ref{fig:ACT+HC_La} for HTPs of YIr$_{2}$Si$_{2}$ and LaIr$_{2}$Si$_{2}$,
respectively, which were found superconducting. Only curves in selected
magnetic fields are displayed from the heat capacity measurements
for better clarity. It is also important to note that we have always
found a sharp drop of resistivity at the SC transition for all the
applied magnetic fields for the YIr$_{2}$Si$_{2}$. The broadened
transitions were observed for LaIr$_{2}$Si$_{2}$ both in zero magnetic
field ($\Delta T=\unit[0.4]{K}$) and also under the applied magnetic
field when two step like transitions developed. 

\begin{figure}
\includegraphics{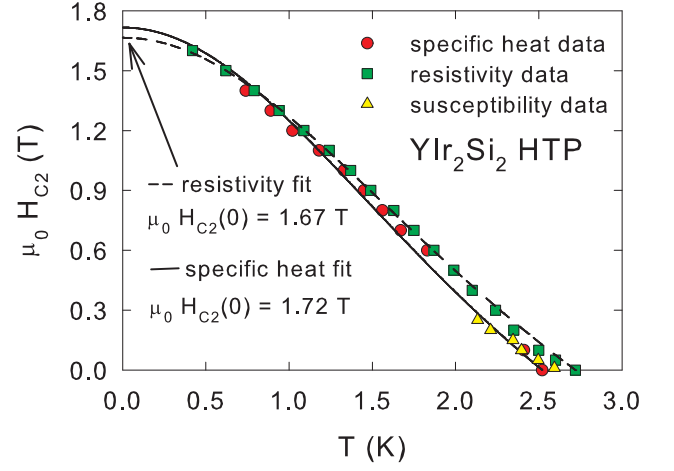}

\caption{\label{fig:HC0_Y}Temperature dependence of the critical fields of
the YIr$_{2}$Si$_{2}$ HTP. Measured data are fitted using Equation
(\ref{eq:GLt}).}

\end{figure}

\begin{figure}
\includegraphics{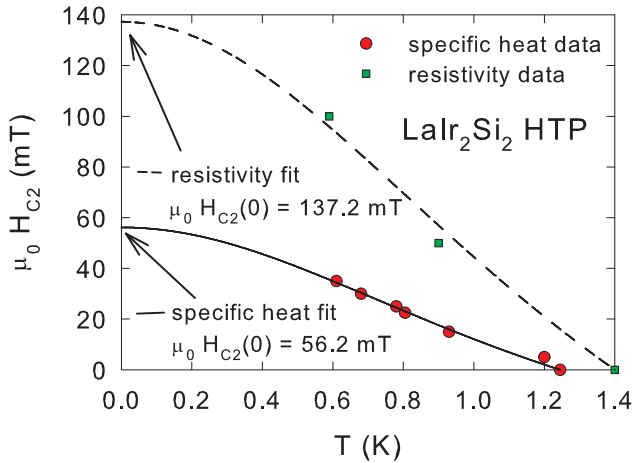}

\caption{\label{fig:HC0_La}Temperature dependence of the critical field of
the LaIr$_{2}$Si$_{2}$ HTP. Measured data are fitted using Equation
(\ref{eq:GLt}).}

\end{figure}

We have plotted all the values of the critical fields from all used
methods and their temperature evolution for both YIr$_{2}$Si$_{2}$
and LaIr$_{2}$Si$_{2}$ in the Figure \ref{fig:HC0_Y} and Figure
\ref{fig:HC0_La}. In the case of YIr$_{2}$Si$_{2}$ we have found
good agreement between the temperature evolution between the resistivity
and heat capacity data. Only the AC susceptibility data measured for
a narrow temperature interval (2.6-2.1 K) are slightly shifted to
lower fields. First we have tested to evaluate the temperature evolution
of the critical field using the square law (Equation (\ref{eq:square}))
.

\begin{equation}
\mu_{0}H_{\mathrm{C2}}\left(T\right)=\mu_{0}H_{\mathrm{C}2}\left(0\right)\left[1-\left(T/T_{\mathrm{SC}}\right)^{2}\right]\label{eq:square}
\end{equation}

The square law was not found as a reasonable model for significant
undervalue of the $\mu_{0}H_{\mathrm{C}2}\left(0\right)$ parameter
in the case of YIr$_{2}$Si$_{2}$. The value of $\mu_{0}H_{\mathrm{C}2}\left(0\right)$
was found to be 1.27 T using square law. We have obtained more satisfying
results using formula derived from Ginzburg-Landau theory \cite{111,112,113}
(Equation (\ref{eq:GLt})).

\begin{equation}
\mu_{0}H_{\mathrm{C2}}\left(T\right)=\mu_{0}H_{\mathrm{C}2}\left(0\right)\left[\frac{1-\left(T/T_{\mathrm{SC}}\right)^{2}}{1+\left(T/T_{\mathrm{SC}}\right)^{2}}\right]\label{eq:GLt}
\end{equation}

This formula gives $\mu_{0}H_{\mathrm{C}2}\left(0\right)=\unit[1.67]{T}$
as a result of the explanation of the resistivity data and similar
value $\mu_{0}H_{\mathrm{C}2}\left(0\right)=\unit[1.72]{T}$ for the
heat capacity data. The value of the critical field for LaIr$_{2}$Si$_{2}$
compound have been also unsuccessfully approximated by the square
law (Equation (\ref{eq:square})) that gives the value of the critical
field $\mu_{0}H_{\mathrm{C}2}\left(0\right)=\unit[42.1]{mT}$ when
applied on the heat capacity data. Also in this case Ginzburg-Landau
formula (Equation (\ref{eq:GLt})) gives better results and estimates
the value of the critical field $\mu_{0}H_{\mathrm{C}2}\left(0\right)=\unit[56.2]{mT}$
for the heat capacity data. Fitting of the resistivity data gives
significantly a higher value of the critical field $\mu_{0}H_{\mathrm{C}2}\left(0\right)=\unit[137.2]{mT}$.
Such behavior unambiguously denoted that the LaIr$_{2}$Si$_{2}$
sample was not in full bulk superconducting state as it will be also
later confirmed by anomaly low value of the $\left(\Delta C_{\mathrm{P}}\right)_{T_{\mathrm{SC}}}/\gamma T_{\mathrm{SC}}$.

On the basis of these findings we have tried to estimate the value
of the $\mu_{0}H_{\mathrm{C}2}\left(0\right)$ from the specific heat
data using Werthamer-Helfand-Hohenberg (WHH) formula (Equation (\ref{eq:whh}))\cite{110}
within the weak-coupling BCS theory, as well. 

\begin{equation}
\mu_{0}H_{\mathrm{C}2}\left(0\right)=-0.693\left(\mathrm{d}H_{\mathrm{C2}}/\mathrm{d}T_{\mathrm{SC}}\right)T_{\mathrm{SC}}\label{eq:whh}
\end{equation}

We have approximated the value of the critical field $\mu_{0}H_{\mathrm{C}2}\left(0\right)\approx\unit[1.6]{T}$
YIr$_{2}$Si$_{2}$ and only $\approx\unit[40]{mT}$ for LaIr$_{2}$Si$_{2}$.
Mainly the WHH value of $\mu_{0}H_{\mathrm{C}2}\left(0\right)$ is
in very good experimental agreement in the case of the Y compound.
All the $\mu_{0}H_{\mathrm{C}2}\left(0\right)$ values are also digestedly
compared and summarized in the Table \ref{tab:summ}.

The specific heat jump at temperature $T_{\mathrm{SC}}=\unit[2.52]{K}$
in zero magnetic field reaches the value $\Delta C_{\mathrm{P}}=\unit[27.9]{mJ\cdot mol^{-1}\cdot K^{-1}}$
in the case of YIr$_{2}$Si$_{2}$ while significantly lower and broadened
jump of $\Delta C_{\mathrm{P}}=\unit[6.7]{mJ\cdot mol^{-1}\cdot K^{-1}}$
occurred for LaIr$_{2}$Si$_{2}$. We have tried to estimate the value
of the $\left(\Delta C_{\mathrm{P}}\right)_{T_{\mathrm{SC}}}/\gamma T_{\mathrm{SC}}$,
which should be 1.43 based on BSC weak-coupling theory. The value
of 1.38 obtained for YIr$_{2}$Si$_{2}$, in good agreement with the
predicted value of 1.43. Contrary to that the corresponding value
for the LaIr$_{2}$Si$_{2}$ is significantly lower than expected
- only 0.67. As we have mentioned hereinbefore we presume that the
sample did not reach the superconducting state at a certain sharp
temperature interval because of broadening of the superconducting
transition probably due to mechanical stresses. Moreover only part
of the sample was superconducting. Similar situation can occur in
intermetallics as was found for example in Ref.\cite{103} because
standard BCS behavior of the La compound was found in Ref.\cite{1}
with $\left(\Delta C_{\mathrm{P}}\right)_{T_{\mathrm{SC}}}/\gamma T_{\mathrm{SC}}=1.3$
which is 90 \% of the BCS-predicted value.

\subsection{First principles electronic structure calculations}

The first principles calculations started minimizing the forces at
the symmetry free Wyckoff positions for both HTP and LTP structures.
We have found very good agreement with our experimental data. Using
fixed spin moment method it has been confirmed Pauli paramagnet as
the ground state of the YIr$_{2}$Si$_{2}$ for both the HTP and LTP.

We now compare the performance of LSDA and GGA with respect to the
equilibrium volume of the HTP and LTP. The ratio \emph{c/a} has been
also minimized and found to be close to the experimental equilibrium
(see Table \ref{tab:c/a-ratios} and \ref{tab:Atomic-coordinates}).
We have calculated the variation of the total energy with the relative
volume $V/V_{0}$ ($V_{0}$ is the experimental equilibrium volume).
The LSDA\cite{61} value of the equilibrium volume is less than 1.0
\% smaller than the experimental value. The GGA from Ref.\cite{62},
on the other hand, leads to a volume that exceeds the experimental
$V_{0}$ by 3.8 \% and the volume obtained with the GGA from Ref.\cite{64}
is 1.3 \% larger. The best results are obtained using the GGA from
Ref.\cite{99}, which underestimates $V_{0}$ by only 1.1\%. An all
forms of the GGA\cite{64,62,99} provide a worse equilibrium volume
than the LSDA. Finally we have calculated the structural difference
energy using LSDA. We have found the value 9 mRy per formula unit
and the LT phase is lower in energy with agreement to our experimental
data. 

\begin{table}
\begin{tabular}{cccc}
\toprule 
 & \multicolumn{2}{c}{YIr$_{2}$Si$_{2}$} & reference\tabularnewline
\midrule
\midrule 
 & LTP & HTP & \tabularnewline
$\left(c/a\right)_{\mathrm{theor}}$ & 2.4474 & 2.3791 & this work\tabularnewline
$\left(c/a\right)_{\mathrm{exp}}$ & 2.4245 & 2.3659 & this work\tabularnewline
\midrule
\midrule 
 & \multicolumn{2}{c}{LaIr$_{2}$Si$_{2}$ } & \tabularnewline
\midrule
\midrule 
 & LTP & HTP & \tabularnewline
$\left(c/a\right)_{\mathrm{theor}}$ & 2.5070 & 2.3749 & Ref.\cite{2}\tabularnewline
$\left(c/a\right)_{\mathrm{exp}}$ & 2.5054 & 2.3734 & this work\tabularnewline
\bottomrule
\end{tabular}

\caption{\emph{\label{tab:c/a-ratios}c/a} ratios as a result of DFT calculations
in comparison with measured data.}

\end{table}

\begin{table}
\begin{tabular}{cccccccc}
\toprule 
\multicolumn{5}{c}{YIr$_{2}$Si$_{2}$ - LTP - experimental data} & \multicolumn{3}{c}{calculated values}\tabularnewline
\midrule
\midrule 
Atom & Symmetry & x & y & z & x & y & z\tabularnewline
\midrule
\midrule 
Si & 4e & 0 & 0 & 0.36409 & 0 & 0 & 0.37934\tabularnewline
Ir & 4d & 0 & 0.5 & 0.25 & 0 & 0.5 & 0.25\tabularnewline
Y & 2a & 0 & 0 & 0 & 0 & 0 & 0\tabularnewline
\midrule
\midrule 
\multicolumn{5}{c}{YIr$_{2}$Si$_{2}$ - HTP - experimental data} & \multicolumn{3}{c}{calculated values}\tabularnewline
\midrule
\midrule 
Atom & Symmetry & x & y & z & x & y & z\tabularnewline
\midrule
\midrule 
Ir & 2c & 0.25 & 0.25 & 0.12753 & 0.25 & 0.25 & 0.12822\tabularnewline
Si & 2c & 0.25 & 0.25 & 0.42296 & 0.25 & 0.25 & 0.36808\tabularnewline
Y & 2a & 0.25 & 0.25 & 0.74891 & 0.25 & 0.25 & 0.7538\tabularnewline
Ir & 2b & 0.75 & 0.25 & 0.5 & 0.75 & 0.25 & 0.5\tabularnewline
Si & 2c & 0.75 & 0.25 & 0 & 0.75 & 0.25 & 0\tabularnewline
\bottomrule
\end{tabular}

\caption{\label{tab:Atomic-coordinates}Atomic coordinates obtained from DFT
calculations compared with measured values.}

\end{table}

We have checked the calculated density of the electronic states and
it is in perfect agreement with the results in Ref.\cite{100}. The
DOS value at the Fermi level is $N(E_{\mathrm{F}})=\unit[2.36]{states/eV\, f.u.}$
and $N(E_{\mathrm{F}})=\unit[2.05]{states/eV\, f.u.}$ which correspond
to $\gamma_{\mathrm{band}}=\unit[5.64]{mJ\cdot mol^{-1}\cdot K^{-2}}$
and $\gamma_{\mathrm{band}}=\unit[4.9]{mJ\cdot mol^{-1}\cdot K^{-2}}$
for the HTP and LTP, respectively. The experimental specific-heat
value is $\gamma_{\mathrm{exp}}=\unit[8.0]{mJ\cdot mol^{-1}\cdot K^{-2}}$
for the HTP, which points to an enhancement factor of $\lambda=0.42$,
$\lambda$ being defined by $\gamma_{\mathrm{exp}}=\gamma_{\mathrm{band}}\left(1+\lambda\right)$.
This total enhancement is most likely due to the electron-phonon coupling.

The first superconducting temperature $T_{\mathrm{SC}}$-relation
based on the minimum set of three parameters (averaged Debye temperature
$\theta_{\mathrm{D}}$, mass-enhancement coefficient $\lambda$ and
a Coulomb pseudopotential $\mu^{*}$) which found extensive applications
in the analysis of superconductors was worked out by McMillan\cite{59}.
Starting with the full Eliashberg equations, McMillan introduced ad
hoc assumptions on the nature of the spectral function and assumed
further that $T_{\mathrm{SC}}$ depends on spectral function only
through $\lambda$. Performing numerical solutions of the Eliashberg
equations McMillan derived so-called McMillan-formula (Equation (\ref{eq:Tsc})).

\begin{equation}
T_{\mathrm{SC}}=\frac{\theta_{\mathrm{D}}}{1.45}\exp\left(-\frac{1.04\left(1+\lambda\right)}{\lambda-\mu^{*}\left(1+0.62\lambda\right)}\right)\label{eq:Tsc}
\end{equation}

Using the averaged Debye temperature $\theta_{\mathrm{D}}=\unit[145]{K}$,
the mass-enhancement coefficient $\lambda=0.42$ and the Coulomb pseudopotential
$\mu^{*}=0.13$ we have found $T_{\mathrm{SC}}=\unit[0.45]{K}$. This
only semiquantitatively agrees with our experimental value $T_{\mathrm{SC}}=\unit[2.52]{K}$.
We note that the result of using McMillan formula is especially very
sensitive to particular value of mass-enhancement coefficient $\lambda$
which is the result of combined analysis of experimental specific
heat data and first-principles calculations based on the DFT with
approximate exchange correlation functional. For example the value
of $\lambda=0.6$ $T_{\mathrm{SC}}\sim\unit[1]{K}$ which is quite
close to our experimental value $T_{\mathrm{SC}}=\unit[2.52]{K}$.
We also point out that we used Coulomb pseudopotential $\mu^{*}=0.13$
which is a common practice to follow suggestion of McMillan\cite{59}
for all transition metals and their compounds. Therefore our calculations
using McMillan formula can be taken as a starting crude estimate of
$T_{\mathrm{SC}}$ only. One possible reason of limiting validity
of using of McMillan formula for our YIr$_{2}$Si$_{2}$ might be
the complex nature of the phonon spectra in YIr$_{2}$Si$_{2}$ compound
and therefore the coupling of electrons to special phonon modes. Therefore
the full first-principles calculation of the superconducting temperature
$T_{\mathrm{SC}}$ of YIr$_{2}$Si$_{2}$ is clearly beyond the scope
of the present paper.

\subsection{Magnetization data}

The experimental magnetic susceptibility of YIr$_{2}$Si$_{2}$ in
the normal state is temperature independent and the value is close
to $\unit[2.5\cdot10^{-8}]{m^{3}\cdot mol^{-1}}$ for HTP and $\unit[1.5\cdot10^{-8}]{m^{3}\cdot mol^{-1}}$
for LTP, respectively. The theoretical value of $\unit[8.4\cdot10^{-8}]{m^{3}\cdot mol^{-1}}$
for HTP and $\unit[7.3\cdot10^{-8}]{m^{3}\cdot mol^{-1}}$ for LTP,
respectively calculated by using the well-known equation for Pauli
susceptibility (Equation (\ref{eq:chi})) provides only right order
with the experimental value. The overestimation of the theoretical
value is probably connected with approximate exchange correlation
functional used in our first principles DFT calculations. 

\begin{equation}
\chi=\mu_{\mathrm{B}}^{2}N\left(E_{\mathrm{F}}\right)\label{eq:chi}
\end{equation}

\begin{table}
\begin{tabular}{ccccc}
\toprule 
parameter & YIr$_{2}$Si$_{2}$ & LaIr$_{2}$Si$_{2}$ & YIr$_{2}$Si$_{2}$ & LaIr$_{2}$Si$_{2}$\tabularnewline
 & HTP & HTP & LTP & LTP\tabularnewline
\midrule
\midrule 
$\unit[T_{\mathrm{SC}}C_{\mathrm{P}}]{\left(K\right)}$ & 2.52 & 1.24 & No SC & No SC\tabularnewline
$\unit[T_{\mathrm{SC}}R]{\left(K\right)}$ & 2.72K & 1.45 & No SC & No SC\tabularnewline
$\unit[T_{\mathrm{SC}}AC\,\chi]{\left(K\right)}$ & 2.7 \textendash{} 2.2 & - & No SC & No SC\tabularnewline
$\unit[T_{\mathrm{SC}}\mathrm{theor.}]{\left(K\right)}$ & 0.45 & - & - & -\tabularnewline
$\unit[\chi_{\mathrm{exp}}]{\left(10^{-8}m^{3}\cdot mol^{-1}\right)}$ & 2.5 & - & 1.5 & 1.8\tabularnewline
$\unit[\chi_{\mathrm{theor}}]{\left(10^{-8}m^{3}\cdot mol^{-1}\right)}$ & 8.4 & - & 7.3 & -\tabularnewline
$\unit[\mu_{0}H_{\mathrm{C2}}\left(0\right)\,\mathrm{WHH}]{\left(T\right)}$ & 1.59 & 0.041 & - & -\tabularnewline
$\unit[\mu_{0}H_{\mathrm{C2}}\left(0\right)\,\mathrm{sq.law}]{\left(T\right)}$ & 1.27 & 0.042 & - & -\tabularnewline
$\unit[\mu_{0}H_{\mathrm{C2}}\left(0\right)\,\mathrm{GL-HC}]{\left(T\right)}$ & 1.72 & 0.056 & - & -\tabularnewline
$\unit[\mu_{0}H_{\mathrm{C2}}\left(0\right)\,\mathrm{GL-R}]{\left(T\right)}$ & 1.67 & 0.137 & - & -\tabularnewline
$\unit[\gamma_{\mathrm{exp}}]{\left(mJ\cdot mol^{-1}\cdot K^{-2}\right)}$ & 8 & 8 & 8.1 & 8.1\tabularnewline
$\unit[\gamma_{\mathrm{theor}}]{\left(mJ\cdot mol^{-1}\cdot K^{-2}\right)}$ & 5.6 & - & 4.9 & -\tabularnewline
$\unit[\xi\left(0\right)\,\mathrm{GL}]{\left(nm\right)}$ & 509 & 88.4 & - & -\tabularnewline
$\unit[\xi\left(0\right)\,\mathrm{WHH}]{\left(nm\right)}$ & 455 & 89.7 & - & -\tabularnewline
$\Delta C/\gamma T_{\mathrm{SC}}$ & 1.38 & 0.67 & - & -\tabularnewline
$\unit[\theta_{\mathrm{D}}]{\left(K\right)}$ & 145 & 155 & 180 & 200\tabularnewline
$\lambda$ & 0.42 & - & - & -\tabularnewline
$\left(c/a\right)_{\mathrm{theor}}$ & 2.3791 & 2.3749\cite{2} & 2.4474 & 2.5070\cite{2}\tabularnewline
$\left(c/a\right)_{\mathrm{exp}}$ & 2.3659 & 2.3734 & 2.4245 & 2.5054\tabularnewline
\bottomrule
\end{tabular}

\caption{\label{tab:summ}List of the parameters of the polymorphs of YIr$_{2}$Si$_{2}$
and LaIr$_{2}$Si$_{2}$.}

\end{table}

\section{conclusions}

We have successfully verified, that polymorphism is an effective tool
to study the influence of the crystal symmetry as a deciding parameter
for presence of the superconductivity on materials. As a model example
we have successfully synthesized both tetragonal polymorphs of the
YIr$_{2}$Si$_{2}$ and LaIr$_{2}$Si$_{2}$ compounds, respectively.
We have determined the role of thermal treatment in formation of the
LTP and the metastable HTP in detail. We conclude that not only the
maximum temperature but also the cooling rate have been found as crucial
parameters to obtain the single-phase material which can explain the
controversial reports on the existence of the superconductivity in
\emph{122} iridium silicides presented in previous works. We have
confirmed that only the HTP is superconducting; $T_{\mathrm{SC}}=\unit[2.52]{K}$
and $\unit[1.24]{K}$ for YIr$_{2}$Si$_{2}$ and LaIr$_{2}$Si$_{2}$,
respectively. The LTP of both, YIr$_{2}$Si$_{2}$ and LaIr$_{2}$Si$_{2}$
, behave as the normal metallic conductors (no sign of superconductivity).
The presence of the superconductivity in Y and La iridium silicides
is connected with unique inverse stacking of Ir-Si layers in the HTP
structure creating pyramidal Ir cages with La or Y atoms inside, which
are not present in LTP polymorphs.

To well understand more physics of the \emph{122} iridium silicides
we have determined all the structure parameters, electronic structure
and we have also inspected the superconducting state of both superconducting
compounds within the BCS theory predictions. 

The lattice parameters and atoms fractional coordinates were found
to be in good agreement with previously published structure models
\cite{4,30} and also in agreement with our theoretical DFT values
obtained using GGA. The LT polymorphs have been also confirmed being
thermodynamically favorable at room temperature. The low values of
the coefficient $\gamma=\unit[8.1]{mJ\cdot mol^{-1}\cdot K^{-2}}$
for the YIr$_{2}$Si$_{2}$ and $\gamma=\unit[8.0]{mJ\cdot mol^{-1}\cdot K^{-2}}$
for the LaIr$_{2}$Si$_{2}$ of the HT polymorph denotes the low density
of states at Fermi level, which was confirmed by theoretical calculations.
We have also revealed a low mass enhancement ($\lambda=0.42$) for
the YIr$_{2}$Si$_{2}$, which indicates a weak electron-phonon interaction.
According to the straightforward use of McMillan formula the theoretical
value of the superconducting temperature is $\unit[0.45]{K}$ which
is not so bad results in comparison to other Y intermetallics. Therefore
we tentatively propose that the observed superconductivity might for
example result from the coupling of electrons to special phonon modes
in the Y and La iridium silicides complex phonon spectra. All physical
parameters and found constants for the YIr$_{2}$Si$_{2}$ and LaIr$_{2}$Si$_{2}$
are digestedly summarized in the table \ref{tab:summ}. 
\begin{acknowledgments}
The work is a part of activities of the Charles University Research
Center \textquotedbl{}Physics of Condensed Matter and Functional Materials.
It was also supported by the Czech Science Foundation (Project \#
202/09/1027) and the Charles University Grant Agency Project \# 719612.
\end{acknowledgments}
\end{document}